\documentclass[]{mn2e}
\usepackage{graphicx,bm,url,times}
\graphicspath{{./fig/}{./png/}}
\newcommand{\EQ}{\begin{equation}}
\newcommand{\EN}{\end{equation}}
\newcommand{\EQA}{\begin{eqnarray}}
\newcommand{\ENA}{\end{eqnarray}}
\newcommand{\eq}[1]{(\ref{#1})}

\newcommand{\Eq}[1]{equation~(\ref{#1})}
\newcommand{\Eqs}[2]{equations~(\ref{#1}) and~(\ref{#2})}

\newcommand{\App}[1]{Appendix~\ref{#1}}
\newcommand{\Sec}[1]{\S\ref{#1}}

\newcommand{\Fig}[1]{Fig.~\ref{#1}}

\newcommand{\bra}[1]{\langle #1\rangle}

{}
\newcommand{\meanFFFF}{\overline{\mbox{\boldmath ${\cal F}$}}{}}{}

{}
{}
{}
{}
{}
{}
{}
{}
{}
{}
{}
{}
{}
{}
\newcommand{\meanUU}{\overline{\mbox{\boldmath $U$}}{}}{}
{}
{}

\newcommand{\meanF}{\overline{\cal F}}
\newcommand{\meanC}{\overline{C}}

\newcommand{\meanFFF}{\overline{\cal F}}
{}
\newcommand{\hatBB}{\hat{\bm{B}}}
{}
\newcommand{\hatOO}{\hat{\bm{\Omega}}}
{}
{}

\newcommand{\Oh}{\hat{\Omega}}

\newcommand{\Bh}{\hat{B}}
\newcommand{\nullvector}{{\bf0}}

\newcommand{\kk}{\bm{k}}
\newcommand{\xx}{\bm{x}}

\newcommand{\uu}{\mbox{\boldmath $u$} {}}
\newcommand{\UU}{\mbox{\boldmath $U$} {}}

\newcommand{\BB}{\mbox{\boldmath $B$} {}}

\newcommand{\JJ}{\mbox{\boldmath $J$} {}}
\newcommand{\SSS}{\mbox{\boldmath $S$} {}}
\newcommand{\AAA}{\mbox{\boldmath $A$} {}}

\newcommand{\ff}{\mbox{\boldmath $f$} {}}

\newcommand{\FF}{\mbox{\boldmath $F$} {}}

\newcommand{\nab}{\mbox{\boldmath $\nabla$} {}}
\newcommand{\OO}{\bm{\Omega}}

\newcommand{\LLLL}{\mbox{\boldmath ${\sf L}$} {}}

\newcommand{\SSSS}{\mbox{\boldmath ${\sf S}$} {}}

\newcommand{\ii}{{\rm i}}

\newcommand{\DD}{{\rm D} {}}

\def\la{\mathrel{\mathchoice {\vcenter{\offinterlineskip\halign{\hfil
$\displaystyle##$\hfil\cr<\cr\sim\cr}}}
{\vcenter{\offinterlineskip\halign{\hfil$\textstyle##$\hfil\cr<\cr\sim\cr}}}
{\vcenter{\offinterlineskip\halign{\hfil$\scriptstyle##$\hfil\cr<\cr\sim\cr}}}
{\vcenter{\offinterlineskip\halign{\hfil$\scriptscriptstyle##$\hfil\cr<\cr\sim\cr}}}}}
\def\ga{\mathrel{\mathchoice {\vcenter{\offinterlineskip\halign{\hfil
$\displaystyle##$\hfil\cr>\cr\sim\cr}}}
{\vcenter{\offinterlineskip\halign{\hfil$\textstyle##$\hfil\cr>\cr\sim\cr}}}
{\vcenter{\offinterlineskip\halign{\hfil$\scriptstyle##$\hfil\cr>\cr\sim\cr}}}
{\vcenter{\offinterlineskip\halign{\hfil$\scriptscriptstyle##$\hfil\cr>\cr\sim\cr}}}}}
\def\Ma{\mbox{\rm Ma}}
\def\Co{\mbox{\rm Co}}

\def\St{\mbox{\rm St}}

\def\Sc{\mbox{\rm Sc}}

\def\Pe{\mbox{\rm Pe}}
\def\Co{\mbox{\rm Co}}
\def\cs{c_{\rm s}}
\def\kf{k_{\rm f}}
\def\urms{u_{\rm rms}}
\def\kappaBB{\kappa_{\rm BB}}
\def\kappaB{\kappa_{\rm B}}
\def\kappaOO{\kappa_{\Omega\Omega}}
\def\kappaO{\kappa_{\Omega}}
\def\kappat{\kappa_{\rm t}}
\def\kappatz{\kappa_{\rm t0}}

\def\BBeq{|\BB|/B_{\rm eq}}
\def\Beq{B_{\rm eq}}

\def\half{{\textstyle{1\over2}}}

\def\onethird{{\textstyle{1\over3}}}

\newcommand{\yapj}[3]{ #1, {ApJ,} {#2}, #3}

\newcommand{\yan}[3]{ #1, {Astron.\ Nachr.,} {#2}, #3}

\newcommand{\yana}[3]{ #1, {A\&A,} {#2}, #3}

\newcommand{\ygafd}[3]{ #1, {Geophys.\ Astrophys.\ Fluid Dyn.,} {#2}, #3}

\newcommand{\yjfm}[3]{ #1, {J.\ Fluid Mech.,} {#2}, #3}

\newcommand{\ypf}[3]{ #1, {Phys.\ Fluids,} {#2}, #3}

\newcommand{\yjetp}[3]{ #1, {Sov.\ Phys.\ JETP,} {#2}, #3}

\newcommand{\yaraa}[3]{ #1, {ARA\&A,} {#2}, #3}

\newcommand{\ymn}[3]{ #1, {MNRAS,} {#2}, #3}

\newcommand{\ysph}[3]{ #1, {Solar Phys.,} {#2}, #3}

\newcommand{\ypre}[3]{ #1, {Phys.\ Rev.\ E,} {#2}, #3}

\newcommand{\yjour}[4]{ #1, {#2}, {#3}, #4}

\newcommand{\ybook}[3]{ #1, {#2} (#3)}
\newcommand{\yproc}[5]{ #1, in {#3}, ed.\ #4 (#5), #2}

\newcommand{\sapj}[1]{ #1, {ApJ}, submitted}

\title{Turbulent diffusion with rotation or magnetic fields}
\author[A. Brandenburg et al.]
{Axel Brandenburg$^1$\thanks{E-mail: brandenb@nordita.org},
Andreas Svedin$^2$ and Geoffrey M. Vasil$^3$
\vspace{2mm}
\\
$^1$NORDITA, AlbaNova University Center, Roslagstullsbacken 23,
SE - 106 91 Stockholm, Sweden\\
$^2$Astronomy Department, Columbia University, New York 10027, USA\\
$^3$JILA, University of Colorado, Boulder, CO 80309-0440, USA}
\date{Accepted 2009 February 16. Received 2009 February 16;
in original form 2009 January 15}
\volume{395}
\begin{document}
\pagerange{\pageref{firstpage}--\pageref{lastpage}} \pubyear{2009}
\maketitle

\begin{abstract}
The turbulent diffusion tensor describing the evolution of the mean
concentration of a passive scalar is investigated for non-helically
forced turbulence in the presence of rotation or a magnetic field.
With rotation, the Coriolis force causes a sideways deflection of the flux
of mean concentration. Within the magnetohydrodynamics approximation there is no
 analogous effect from the magnetic field because the effects on the flow do not depend on the sign of the field.
Both rotation and magnetic fields tend to suppress turbulent transport,
but this suppression is weaker in the direction along the magnetic field.
Turbulent transport along the rotation axis is not strongly affected by
rotation, except on shorter length scales, i.e., when the scale of the
variation of the mean field becomes comparable with the scale of the
energy-carrying eddies.
These results are discussed in the context of anisotropic convective
energy transport in the Sun.
\end{abstract}
\label{firstpage}
\begin{keywords}
magnetic fields --- MHD --- hydrodynamics -- turbulence
\end{keywords}

\section{Introduction}

In the outer 30\% of the sun, energy is transported mainly by convection
through the exchange of fluid parcels. This process tends to smear out entropy 
gradients by making the gas nearly isentropic on the average.
Any remaining entropy gradients lead to a mean entropy flux whose
negative divergence leads to a local change in the entropy.
This entropy flux is equal to the negative gradient of the mean entropy
multiplied by some turbulent diffusivity coefficient.
Using mixing length theory, this turbulent diffusion coefficient is found
to be of the order of the turbulent velocity times some mixing length,
which is often taken as a free parameter (Vitense 1953).

With rotation, the smearing out of entropy gradients is significantly
reduced due to lateral mixing generated by Coriolis deflections
(Julien et al.\ 1996).  Such deflections cause fluid parcels to lose their 
heat content to the sides before they can ascend or descend and transport heat vertically.  
This additional mixing explains why the smearing out of entropy gradients is heavily reduced by rotation.
Rotational effects on the turbulent diffusivity were already
studied by Weiss (1965) and Durney \& Roxburgh (1971) who found that the
turbulent diffusion tensor becomes anisotropic with respect to
the direction of the rotation axis.   The components of the anisotropic
diffusion tensor can be computed using the first-order smoothing approximation (R\"udiger 1989, Kitchatinov et al.\ 1994).
These calculations show that the enhancement of turbulent diffusion
along the direction of rotation corresponds to a latitudinal heat flux
toward the pole, making it slightly hotter.
Turbulent convection simulations by R\"udiger et al.\ (2005) confirm this,
but they also show that the radial heat flux is slightly smaller near
the poles, which is explained by the effect of stratificational anisotropy.
This agrees with earlier results of Rieutord et al.\ (1994) and
K\"apyl\"a et al. (2004).
R\"udiger et al.\ (2005) also established the presence of an azimuthal
turbulent heat flux and found it to be negative, i.e., in the westward
direction.
This was surprising in view of the result of Kitchatinov et al.\ (1994)
who found it to be either zero (for a simple mixing length model) or
positive (for a turbulence model with finite-time correlation effects).
However, their result was based on a sign error (R\"udiger, private communication).
Furthermore, if there is also stratification, the simple mixing length model
of Kitchatinov et al.\ (1994) predicts that the sign of the flux should
agree with that of the corresponding component of the Reynolds stress
tensor and hence negative, which is then in agreement with the convection simulations of R\"udiger et al.\ (2005).

Diffusive processes also play a role in the evolution of the mean
magnetic field and the mean momentum.
In recent years it has become possible to compute the relevant
diffusion coefficients together with other (e.g.\ non-diffusive)
turbulent transport coefficients using turbulence simulations.
In the case of magnetic fields, a fairly accurate method is the test field method in which one solves an additional set
of passive evolution equations for the departures from the corresponding
set of prescribed mean fields (Schrinner et al.\ 2005, 2007;
Brandenburg 2005, Brandenburg et al.\ 2008a).

In cases of poor scale separation, i.e.\ when the scale of the mean field
is only a few times larger than the typical scale of the turbulence,
the multiplication with a diffusion coefficient must be replaced by
a convolution with an integral kernel.
A similar situation also applies to cases where the timescale of the
variation of the mean fields becomes comparable with the turnover time.
In such cases, the test field method has also been used to determine the integral 
kernels (Brandenburg et al.\ 2008a, Hubbard \& Brandenburg 2008). 
If the turbulence is anisotropic and/or inhomogeneous, the diffusion coefficients or kernels are replaced by tensors.
Again, the test field method allows an accurate determination of all
relevant tensor components.

In this paper, we apply the test field method to the turbulent diffusion of
entropy in cases where the turbulence is homogeneous, but anisotropic
due to the presence of rotation or a magnetic field.
The former problem has recently been addressed by R\"udiger et al.\ (2005),
who evaluated the actual fluxes in stratified turbulent convection. 
However, convection necessarily imposes an additional
anisotropy in the direction of gravity, so it becomes harder to disentangle the effects of gravity and rotation.
Here, we focus on the case of forced turbulence in the absence of gravity; using an isothermal equation of state.
In that case, the evolution equation of entropy is similar to that
of a passive scalar.

The effect of rotation on turbulent diffusion is important for modeling
the possibility of latitudinal entropy gradients in rotating stars, 
i.e.\ perpendicular to the radial direction of
the energy flux.  Latitudinal entropy gradients could be important for producing a 
significant baroclinic term that would balance the curl of the 
Coriolis force and could therefore be important for explaining the observation 
that the contours of constant angular velocity in the Sun are not parallel to the 
rotation axis (Kitchatinov \& R\"udiger 1995; Thompson et al.\ 2003).  The effect of a magnetic
field on turbulent heat diffusion is important in the theory of sunspots (e.g.\ Kitchatinov, \& Mazur 2000)
where the energy transport is suppressed in an anisotropic fashion
(R\"udiger \& Hollerbach 2004). Furthermore, anisotropic passive scalar 
transport is important in connection with understanding the partial lithium depletion in the Sun (R\"udiger \& Pipin 2001).

\section{Preliminary considerations}
\label{Preliminary}

We make use of the fact that in the absence of gravity, and with an
isothermal equation of state the evolution
of entropy is passive, i.e.\ there is no feedback on the momentum equation. 
We can therefore consider the evolution equation for a passive scalar concentration per unit volume $C$, which obeys
\EQ
{\partial C\over\partial t}=-\nab\cdot(\UU C)+\kappa\nabla^2C,
\label{dCdt}
\EN
where $\kappa$ is the diffusivity and $\UU$ is the velocity.
We note that in the compressible case there is an extra term
$-\kappa\nab\cdot(C\nab\ln\rho)$ on the right hand side of \Eq{dCdt}
where $\rho$ is the density.

The evolution of the mean concentration $\meanC$ is obtained by averaging
\Eq{dCdt}, which yields
\EQ
{\partial\meanC\over\partial t}=-\nab\cdot(\meanUU\,\meanC+\overline{\uu c})
+\kappa\nabla^2\meanC,
\label{dmeanCdt}
\EN
where $\overline{\uu c}\equiv\meanFFFF$ is the mean flux of the
concentration density.
Under the assumption that the spatial and temporal scales of the
mean field are well separated from those of the turbulence, one
can make a generalized Fickian diffusion ansatz of the form
\EQ
\meanF_i=-\kappa_{ij}\nabla_j\meanC
\label{meanF}
\EN
where $\kappa_{ij}$ is the turbulent diffusion tensor.
In the absence of good scale separation the multiplication with
a diffusion tensor must be replaced by a convolution with an
integral kernel. We discuss this case near the end of the paper, but ignore this for now.  

The turbulent diffusion tensor is a proper tensor, so in
the presence of rotation and in the absence of helicity
it has only even powers of the
pseudo vector $\OO$ and odd powers of $\OO$ combined with
odd powers of the Levi-Civita tensor.
It must therefore have the general form (Kitchatinov et al.\ 1994)
\EQ
\kappa_{ij}
=\kappat\delta_{ij}
+\kappaO\epsilon_{ijk}\Oh_k
+\kappaOO\Oh_i\Oh_j,
\label{kappaij}
\EN
where $\hatOO=\OO/|\OO|$ is the unit vector along the rotation axis and
$\kappat$, $\kappaO$, and $\kappaOO$ are functions
of the Coriolis, Peclet, and Schmidt numbers, that are defined by
\EQ
\Co={2\Omega\over\urms\kf},\quad\Pe={\urms\over\kappa\kf},
\quad\mbox{and}\quad\Sc={\nu\over\kappa},
\EN
respectively.  Incidentally, we note that $\kappaO$ contributes an anti-symmetric 
component to the turbulent diffusion tensor, which does not contribute to divergence of flux. 
Similarly, in the presence of a magnetic field $\BB$ with unit vector  $\hatBB=\BB/|\BB|$, the diffusion tensor must have the form
\EQ
\kappa_{ij}
=\kappat\delta_{ij}
+\kappaB\epsilon_{ijk}\Bh_k
+\kappaBB\Bh_i \Bh_j,
\EN
but here $\kappaB=0$, because the diffusion tensor can only depend on
the Lorentz force which is quadratic in $\BB$, so $\kappa_{ij}$ must be
invariant under a change of sign of $\BB$ (Kitchatinov et al.\ 1994).
Thus, we have only $\kappat$ and $\kappaBB$ that are functions of
$\BBeq$, $\Pe$, and $\Sc$, where
\EQ
\Beq=\bra{\mu_0\rho\UU^2}^{1/2}
\EN
is the equipartition field strength, and $\mu_0$ is the vacuum permeability.

By adding rotation or by imposing a magnetic field, the flow becomes
anisotropic.  The degree of anisotropy is characterized by the parameter
\EQ
\varepsilon=1-\overline{u_\perp^2}/\overline{u_\parallel^2}
\label{varepsilon}
\EN
where $\overline{u_\perp^2}$ and $\overline{u_\parallel^2}$ are the
velocity dispersions perpendicular and parallel to the direction of
anisotropy, $\hatOO$ or $\hatBB$.
In practice, we take these directions to be the $z$ direction, in which case
$\overline{u_\parallel^2}=\overline{u_z^2}$ and $\overline{u_\perp^2}$
is calculated as $\half(\overline{u_x^2}+\overline{u_y^2})$.
Isotropy implies $\varepsilon=0$.

\section{Method}

We simulate turbulence by solving the compressible hydromagnetic equations
with an imposed random forcing term and an isothermal equation of state,
so that the pressure $p$ is related to $\rho$ via
$p=\rho\cs^2$, where $\cs$ is the isothermal sound speed.
We consider a periodic cartesian domain of size $L^3$.
In the presence of an imposed field $\BB_0$, we write the total field as 
$\BB=\BB_0+\nab\times\AAA$, where $\AAA$ is the magnetic vector potential.
In the nonmagnetic case, we put $\BB_0=\nullvector$, and with $\AAA=\nullvector$
initially, we have $\BB=0$ for all times, and the hydromagnetic equations
reduce to the Navier-Stokes equations for $\rho$ and $\UU$,
\EQ
{\DD\ln\rho\over\DD t}=-\nab\cdot\UU,
\EN
\EQ
{\DD\UU\over\DD t}=-2\OO\times\UU-\cs^2\nab\ln\rho
+\ff+\FF_{\rm visc}+\FF_{\rm Lor},
\label{dUdt}
\EN
where $\FF_{\rm Lor}=\nullvector$ in the non-magnetic case,
i.e.\ there is no Lorentz force,
$\DD/\DD t=\partial/\partial t+\UU\cdot\nab$ is the advective
derivative, $\OO$ is the background angular velocity,
$\FF_{\rm visc}=\rho^{-1}\nab\cdot2\rho\nu\SSSS$ is the viscous
force, $\nu$ is the kinematic viscosity,
$\SSS_{ij}=\half(U_{i,j}+U_{j,i})-\onethird\delta_{ij}\nab\cdot\UU$
is the traceless rate of strain tensor, and $\ff$ is a random forcing
function consisting of plane transversal waves with random wavevectors
$\kk$ such that $|\kk|$ lies in a band around a given forcing wavenumber
$k_{\rm f}$.
The vector $\kk$ changes randomly from one timestep to the next.
This method is described for example in Haugen et al.\ (2004).
The forcing amplitude is chosen such that the Mach number $\Ma=\urms/\cs$
is about 0.1.

In the magnetic case, we have a finite Lorentz force
$\FF_{\rm Lor}=\JJ\times\BB/\rho$, where
$\JJ=\nab\times\BB/\mu_0$ is the current density, $\mu_0$ is the vacuum permeability, 
and we solve the uncurled induction equation in the form
\EQ
{\partial\AAA\over\partial t}=\UU\times\BB+\eta\nabla^2\AAA,
\EN
where $\eta$ is the magnetic diffusivity, which is assumed constant.

The test scalar equation is obtained by subtracting \Eq{dmeanCdt} from
\Eq{dCdt}, which yields
\EQ
{\partial c\over\partial t}=
-\nab\cdot(\meanUU c+\uu\meanC+\uu c-\overline{\uu c})
+\kappa\nabla^2c,
\label{dcdt}
\EN
where $c=C-\meanC$ and $\uu=\UU-\meanUU$ are the fluctuating values of 
concentration and velocity and $\overline{\uu c}$ is the mean flux of concentration density.
We restrict ourselves to the case of two-dimensional mean fields,
$\meanC=\meanC(x,z,t)$, that are obtained by averaging over the
$y$ direction.
In the spirit of the test field method, we solve \Eq{dcdt} for a
predetermined set of four different mean fields,
\EQ
\meanC^{c0}=C_0\cos kx,\quad\meanC^{s0}=C_0\sin kx,
\EN
\EQ
\meanC^{0c}=C_0\cos kz,\quad\meanC^{0s}=C_0\sin kz,
\EN
where $C_0$ is a normalization factor, that will drop out in the 
determination of $\kappa_{ij}$ since \Eq{meanF} is linear.
For each of these mean fields $\meanC^{pq}$ we obtain a separate
evolution equation for $c^{pq}$,
\EQ
{\partial c^{pq}\over\partial t}=
-\nab\cdot(\meanUU c^{pq}+\uu\meanC^{pq}+\uu c^{pq}-\overline{\uu c^{pq}})
+\kappa\nabla^2c^{pq},
\label{dcpqdt}
\EN
where $p,q=c$, $s$, or 0.
In this way , we calculate four different fluxes,
$\meanFFFF^{pq}=\overline{\uu c^{pq}}$, and compute
the 6 relevant components of $\kappa_{ij}$
\EQ
\kappa_{i1}=-\bra{\cos kx\meanFFF_i^{s0}-\sin kx\meanFFF_i^{c0}}/k,
\label{Fx}
\EN
\EQ
\kappa_{i3}=-\bra{\cos kz\meanFFF_i^{0s}-\sin kz\meanFFF_i^{0c}}/k,
\label{Fz}
\EN
for $i=1, ..., 3$.
Here, angular brackets denote volume averages.
The expressions for $\kappa_{i1}$ and $\kappa_{i3}$ in \Eqs{Fx}{Fz} 
differ in that the superscripts $s0$ and $c0$ are replaced by $0s$ and $0c$, respectively.
Note that the components of $\kappa_{ij}$ for $j=2$ are irrelevant for mean fields that are independent of $y$.

The method provides the functions $\kappat$, $\kappaO$, and $\kappaOO$
(or $\kappaB$ and $\kappaBB$) as functions of $z$ and $t$.
Unless otherwise specified, we quote averages over $z$ and $t$.
Error bars are calculated by averaging the results first over $z$ and
then over subsequent time intervals $\Delta t$ whose length corresponds
to one correlation time, i.e.\ $\Delta t\,\urms\kf=1$.
This results in a number of independent measurements from which we
calculate the standard deviation, which is what we quote as the error.

The simulations have been carried out using the {\sc Pencil
Code}\footnote{\url{http://www.nordita.org/software/pencil-code}} which
is a high-order finite-difference code (sixth order in space and third
order in time) for solving the compressible hydromagnetic equations.
As of revision {\tt r10245}, the test scalar equations are
implemented into the public-domain code similar to the way described
above and are invoked by compiling with {\tt TESTSCALAR=testscalar}
\footnote{\url{http://pencil-code.googlecode.com/}}.
The numerical resolution used in the simulations depends on the Peclet
number and reaches $256^3$ meshpoints for our runs with $\Pe\approx200$.

\section{Results}

As in earlier work on determining the turbulent magnetic diffusivity
using the test field method,
we present our results for the relevant components of $\kappa_{ij}$
by normalizing them with respect to the reference value
\EQ
\kappa_{\rm t0}=\onethird\urms\kf^{-1}.
\EN
This is motivated by the first order smoothing result
$\kappat=\onethird\tau\urms^2$ and the assumption that the Strouhal number
$\St=\tau\urms\kf$ is about unity (Brandenburg et al.\ 2004).
We focus on the case $\kf/k_1\approx3$, which is a compromise
of having enough wavenumbers between $\kf$ and the dissipation wavenumber
(which is where an inertial range can develop), and leaving some room for
scales larger than that of the energy-carrying eddies corresponding to
wavenumber $\kf$.
On a few occasions we also consider other values of $\kf/k_1$.

\begin{figure}\begin{center}
\includegraphics[width=\columnwidth]{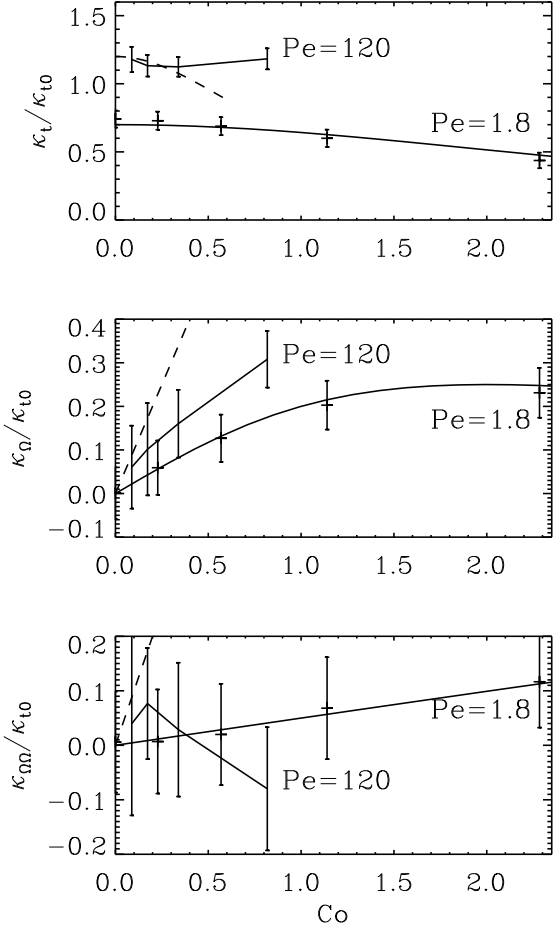}
\end{center}\caption[]{
$\Co$ dependence of $\kappat$, $\kappaO$, and $\kappaOO$
for $\Pe=1.8$ and $120$.
For $\Pe=1.8$ the results can be fitted to expressions described in the text,
while for larger values of $\Pe$ the dependences tend to be more complicated.
The dashed lines indicate the behavior expected from the simple treatment of
the $\tau$ approximation in \App{KappaOmega}.
The slope of the dashed lines in the middle and last panel is therefore 1.
For these runs the anisotropy parameter $\varepsilon$ varies between 0 and 0.04.
}\label{p16b}\end{figure}

We begin by considering the case with rotation and no magnetic field.
In \Fig{p16b} we compare the dependence of $\kappat$,
$\kappaO$, and $\kappaOO$ on $\Co$ for
fixed values of $\Pe=1.8$ and $120$, using $\Sc=1$ in all cases.
Note that $\kappat$ shows a mild decline with $\Co$ while both
$\kappaO$ and $\kappaOO$ are positive and increase
with $\Co$.
For $\Pe=1.8$ the results can be fitted to expressions of the form
$\kappat\approx0.7/[1+(0.3\,\Co)^2]$,
$\kappaO\approx0.25\,\Co/[1+(0.5\,\Co)^2]$, and
$\kappaOO\approx0.05\,\Co/[1+(0.05\,\Co)^2]$.
For larger values of $\Pe$ the dependence of $\kappat$ and
$\kappaO$ on $\Co$ is qualitatively similar, but the dependence
of $\kappaOO$ on $\Co$ now shows
indications of a sign change for larger values of $\Co$.
A positive $\kappaO$ would agree with the result of
R\"udiger et al.\ (2005), although their result was explained to be
due to the additional effect of stratification.

\begin{figure}\begin{center}
\includegraphics[width=\columnwidth]{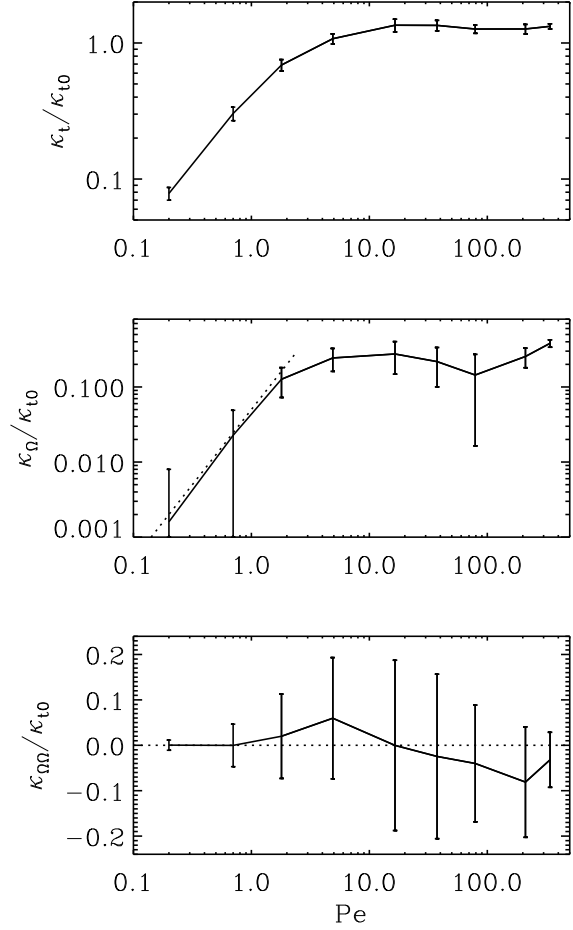}
\end{center}\caption[]{
$\Pe$ dependence for $\Co=0.5$ keeping $\Sc=1$ in all cases.
The resolution is increased with increasing values of $\Pe$ and
reaches $256^3$ for the run with the largest value of $\Pe$.
}\label{pOm01}\end{figure}

The dependence of $\kappat$, $\kappaO$, and $\kappaOO$
on $\Pe$ for $\Co=0.5$ and $\Sc=1$ is shown in \Fig{pOm01}.
As expected, for very small values of $\Pe$ all three functions go to zero like $\kappat/\kappatz\approx0.4\,\Pe$ and
$\kappaO/\kappatz\approx0.05\,\Pe^2$, while $\kappaOO$
is close to zero and, within error bars, essentially compatible with zero.
For large values of $\Pe$, both $\kappat$ and $\kappaO$ seem to reach
asymptotic values, while for $\kappaOO$ there is no clear trend.
An asymptotic value of $\kappat$ was expected based on the analogy with the 
magnetic case where the magnetic diffusivity was also found to reach an asymptotic value for large Reynolds numbers (Sur et al.\ 2008).

\begin{figure}\begin{center}
\includegraphics[width=\columnwidth]{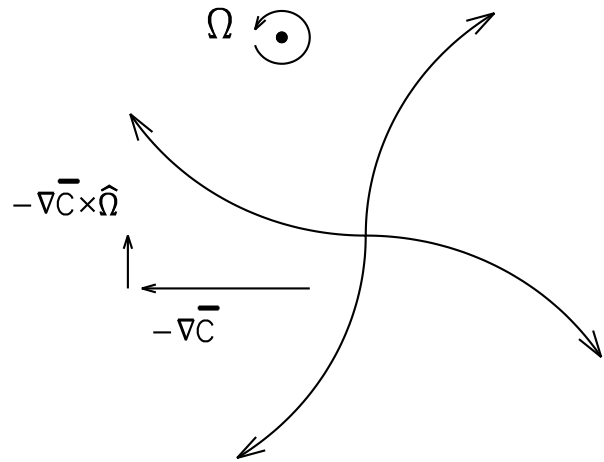}
\end{center}\caption[]{
Sketch illustrating the systematic sideway motion caused by the
Coriolis force and its effect on the flux and hence the sign of $\kappaO$.  
}\label{psketch}\end{figure}

In all our cases we find that $\kappaO$ is positive.
This result is in agreement with that of R\"udiger et al.\ (2005), who
define a quantity $\tilde{\chi}$ such that $\tilde{\chi}\Omega=-\kappaO$.
They point out that in spherical coordinates $(r,\theta,\phi)$ a finite
$\kappaO$ corresponds to a finite $\kappa_{r\phi}$ which, in turn,
leads to a finite azimuthal heat flux. Its astrophysical importance is limited, because the azimuthal average of its divergence vanishes.
However, R\"udiger et al.\ (2005) stress that the sign and magnitude of
$\kappaO$ provides an important test of mean-field theory.
Their result was motivated using a result from the mixing-length model
of Kitchatinov et al.\ (1994) that $\kappa_{ij}$ should be proportional
to the Reynolds-stress tensor $\overline{u_i u_j}$.
A positive $\kappaO$ would correspond to a negative $\kappa_{r\phi}$,
which could be explained by a negative $\overline{u_r u_\phi}$ that is
expected for stratified turbulence with dominant vertical turbulent velocities.
This argument, which is originally due to Gough (1978), does not apply in our case.
However, a positive value of $\kappaO$ is an immediate consequence
of the Coriolis force producing systematic sideway motions;
see \Fig{psketch}.  That is, suppose there exists a mean concentration gradient in, say, the positive $x$-direction, i.e., $\boldsymbol{\nabla}\overline{C} \sim \boldsymbol{\hat{x}}$.  Then a small fluid parcel that is moving in the down-gradient direction, $\boldsymbol{u} \sim -\boldsymbol{\nabla}\overline{C}$, would experience, via the Coriolis acceleration, a deflection into the positive $y$-direction.  The mean effect of such deflections would be to produce a positive coherent flux in the direction given by $- \boldsymbol{\nabla}\overline{C}\times\boldsymbol{\hat{\Omega}}$, and thus $\kappaO $ must be positive.  A positive value of $\kappaO$ can also be explained using the pressureless $\tau$ approximation which predicts
$\kappaO/\kappat=\Co$; see \App{KappaOmega}.
This underlines the potential usefulness of the $\tau$ approximation in
mean-field theory, even if this approach is used in its most rudimentary
form.

The simple treatment of the $\tau$ approximation in \App{KappaOmega}
predicts that $\kappat$ declines with increasing values of $\Co$,
while both $\kappaO$ and $\kappaOO$ increase and are positive.
This is in agreement with the simulations, but quantitative details are
not reproduced: in the simulations the decline of $\kappat$ as well as
the increase of $\kappaO$ and $\kappaOO$ are slower than what is predicted
by the simple treatment of the $\tau$ approximation.
There could be several reasons for this departure of the simulations from theory.
Firstly, the definition of $\Co$ in not unique.
However, even for small values of $\Pe$ the functions
$\kappat$, $\kappaO$, and $\kappaOO$ would require different
scaling factors on $\Co$ to fit the results (see above).
Secondly, the feedback of rotational effects on the Reynolds stresses
are not taken into account in the simple form of the $\tau$ approximation.
On the other hand, the rotational anisotropy parameter $\varepsilon$
remains fairly small (below 0.04) even for the largest values of $\Co$
considered.
The correlation $\overline{\ff c}$ was neglected, but this should be
permissible in the current case with $\delta$-correlated forcing; see
Sur et al.\ (2007) for the corresponding case with magnetic fields.
Finally, the neglect of the pressure term in the simple form of the
$\tau$ approximation is not justified, and a more rigorous treatment
might alter the results significantly.

\begin{figure}\begin{center}
\includegraphics[width=\columnwidth]{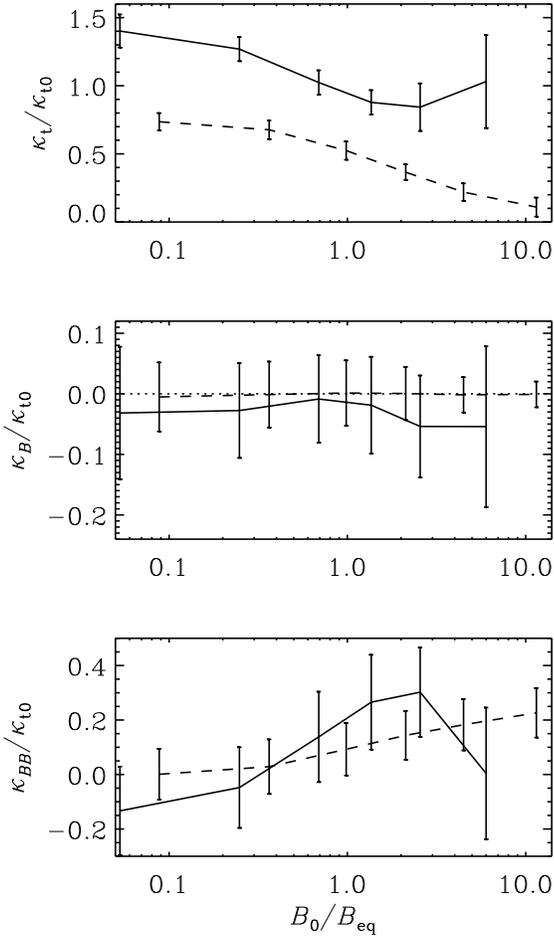}
\end{center}\caption[]{
Dependence on the magnetic field strength for $\Pe=1.8$ (dashed)
and for $\Pe$ in the range 12--15 (solid).
For these runs the anisotropy parameter $\varepsilon$ varies between 0 and 0.04.
}\label{p32c_16a}\end{figure}

Next, we consider the case with a magnetic field and
look at the dependence of $\kappat$ and $\kappaBB$
on the magnetic field strength for fixed Pe=1.8;
see \Fig{p32c_16a}.
The function $\kappaB$ is also plotted, but, as expected,
it is compatible with zero.
Again, $\kappat$ decreases with increasing field strength, but
$\kappaBB$ is finite and positive, and increases with $B_0$.

\begin{figure}\begin{center}
\includegraphics[width=\columnwidth]{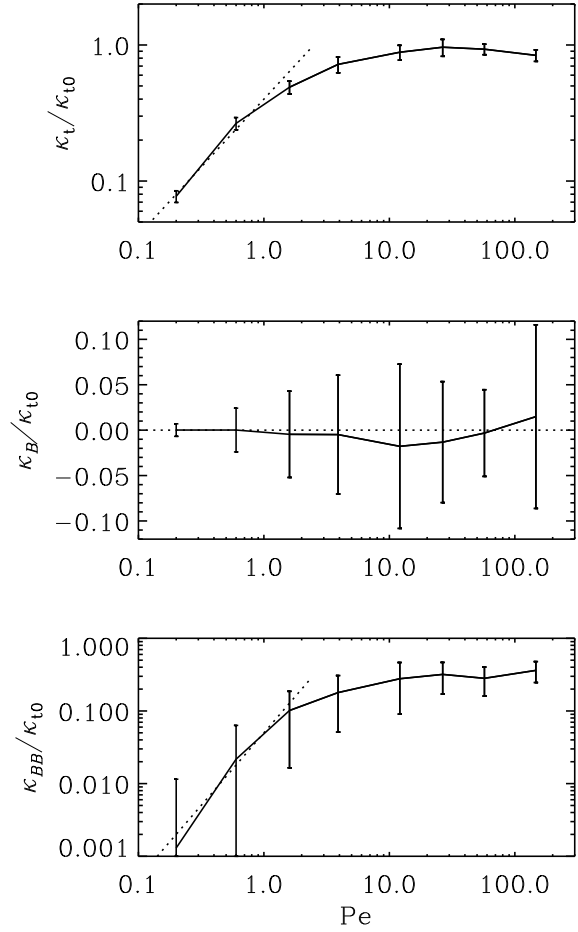}
\end{center}\caption[]{
$\Pe$ dependence for $\Co=0$ and $B_0/B_{\rm eq}=1.3$.
The dotted lines in the first and last panel correspond to
$0.4\,\Pe$ and $0.05\,\Pe^2$, respectively.
For the run with the largest value of $\Pe$ we have $\varepsilon=0.20$.
}\label{pB01}\end{figure}

The dependence of $\kappat$ and $\kappaBB$ on $\Pe$ is shown in \Fig{pB01}.
Both functions increase with $\Pe$ as long as $\Pe$ is below about 10, and
then seem to level off, although it would be desirable to confirm this
for larger values of $\Pe$.
The $\Pe$ dependence of $\kappat$ is similar to that for $B_0=0$
and $\Co=0.6$ where $\kappat$ goes to zero with decreasing $\Pe$ like
$0.4\,\Pe$, and $\kappaBB\approx0.05\,\Pe^2$, which is reminiscent
of the behavior of $\kappaO$ for $B_0=0$ and $\Co=0.6$.

Finally, we consider the dependence of the components of $\kappa_{ij}$
on the wavenumber $k$ of the test scalar in \Eqs{Fx}{Fz}.
A dependence of $\kappa_{ij}$ on $k$ reflects the fact that there is
poor scale separation, i.e.\ $k/\kf$ is not very small.
In such a case (as already discussed in \Sec{Preliminary}), the multiplication
with a turbulent diffusivity in \Eq{meanF} must be replaced by a
convolution with an integral kernel, as in
Brandenburg et al.\ (2008b) for the case of magnetic diffusion and
$\alpha$ effect.  In Fourier space the convolution corresponds to a multiplication.  
As long as we consider monochromatic mean fields 
(proportional to $e^{\ii\kk\cdot\xx}$ for a single wavevector) the values 
of the numerically determined components of $\kappa_{ij}$ correspond to the 
value of the Fourier transform $\tilde\kappa_{ij}$ at wavevector $\kk$. 
The full integral kernel can then be assembled by determining the full 
$\kk$ dependence and then Fourier transforming back into real space.

\begin{figure}\begin{center}
\includegraphics[width=\columnwidth]{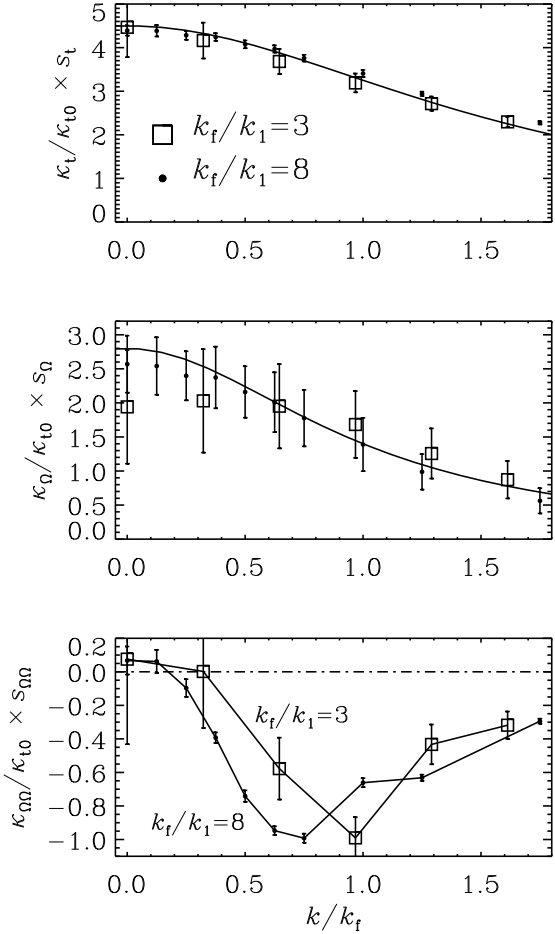}
\end{center}\caption[]{
Wavenumber dependence of $\kappat$, $\kappaO$, and
$\kappaOO$ for $\Co=0.6$ and $\kf/k_1=8$ with $\Pe=70$
compared with the case $\kf/k_1=3$ with $\Pe=17$.
In the upper two panels the lines represent curves of the form
\eq{Lorentzian} with $a=0.62$ and 0.5, respectively.
For these runs the anisotropy parameter $\varepsilon$ is around 0.003
for the runs with $\kf/k_1=8$ and $\Pe=70$ and around 0.036
for the runs with $\kf/k_1=3$ and $\Pe=17$.
}\label{p32d}\end{figure}

In \Fig{p32d} we show the resulting dependences $\tilde\kappat(k)$,
$\tilde\kappaO(k)$, and $\tilde\kappaOO(k)$,
similar to earlier work on turbulent magnetic diffusion (Brandenburg et al.\ 2008b).  
It turns out that both $\tilde\kappat$ and $\tilde\kappaO$
can be fitted reasonably well to a Lorentzian,
\EQ
\tilde\kappa_i={\tilde\kappa_{i0}\over1+(a k/\kf)^2},
\label{Lorentzian}
\EN
where $a\approx0.62$.
This value of $a$ is close to the corresponding value for turbulent
magnetic diffusion where $a\approx0.5$ was found for the isotropic
case (Brandenburg et al.\ 2008b) and $a\approx0.7$ in the anisotropic
case with shear, but no rotation (Mitra et al.\ 2009).
However, the coefficients $\tilde\kappatz$ and $\tilde\kappa_{\Omega0}$
depend on value of $\kf/k_1$.
For $\kf/k_1=3$ we find $\tilde\kappatz/\kappatz=1.4$
and $\tilde\kappa_{\Omega0}\approx0.3$, while
for $\kf/k_1=8$ we find $\tilde\kappatz/\kappatz=1.5$
and $\tilde\kappa_{\Omega0}/\kappatz\approx0.5$.
To make the curves for different values of $\kf$ and $\Pe$ more nearly
overlap, we have scaled $\kappat$ by a factor $s_{\rm t}=\kf/k_1$,
$\kappaO$ by a factor $s_{\Omega}=(\kf/k_1)^{1.7}$, and
$\kappaOO$ by a factor $s_{\Omega}=(\kf/k_1)^{0.5}$.

\begin{figure}\begin{center}
\includegraphics[width=\columnwidth]{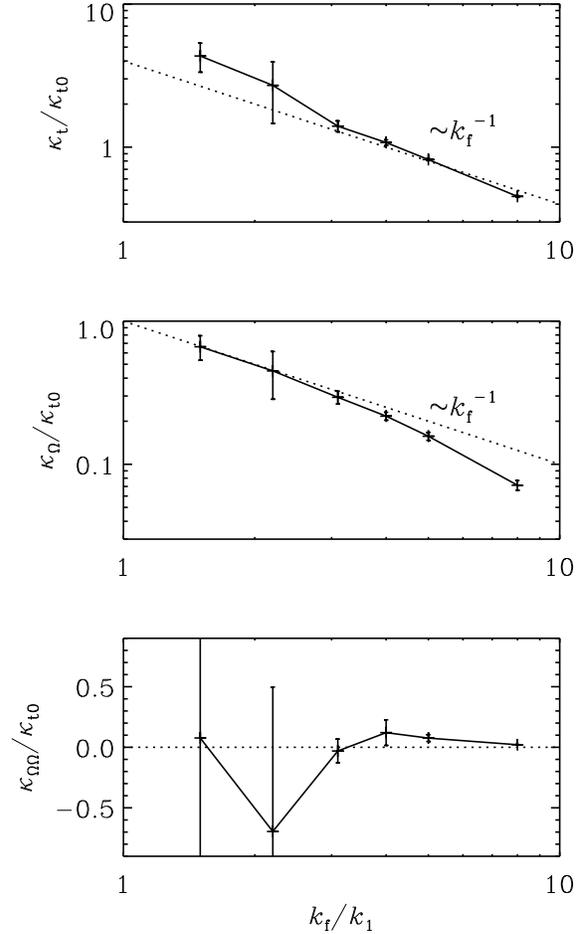}
\end{center}\caption[]{
Dependence of $\kappat$, $\kappaO$, and $\kappaOO$ on the forcing
wavenumber $\kf/k_1$ for $k=0$ with $\Co=0.6$ and $\Pe\approx17$.
Note the scaling of $\kappat$ and $\kappaO$ with $\kf^{-1}$,
while $\kappaOO$ is essentially compatible with zero.
}\label{p32a_kf}\end{figure}

A surprising result occurs for $\tilde\kappaOO(k)$ which takes
large negative values for $k/\kf\approx1$, but tends to vanish
for $k\to0$. This means that the diffusion of mean fields along the 
axis of rotation is strongly suppressed once its typical length scale
becomes comparable with the scale of the energy-carrying eddies.

The two cases shown in \Fig{p32d} differ not only in the value
of $\kf/k_1$ (3 and 8), but also in the value of $\Pe$ (17 and 70).
In order to show that the dependence of the scaling coefficients
$\tilde\kappatz$ and $\tilde\kappa_{\Omega0}$ is primarily due to
the change in $\kf/k_1$ we plot in \Fig{p32a_kf} the dependence of
$\kappat$, $\kappaO$, and $\kappaOO$ on the forcing wavenumber
$\kf/k_1$ for $k=0$ with $\Co=0.6$ and $\Pe\approx17$.
It turns out that both $\kappat$ and $\kappaO$ decrease with $\kf$
like $\kf^{-1}$, while $\kappaOO$ is essentially compatible with zero.

\begin{figure}\begin{center}
\includegraphics[width=\columnwidth]{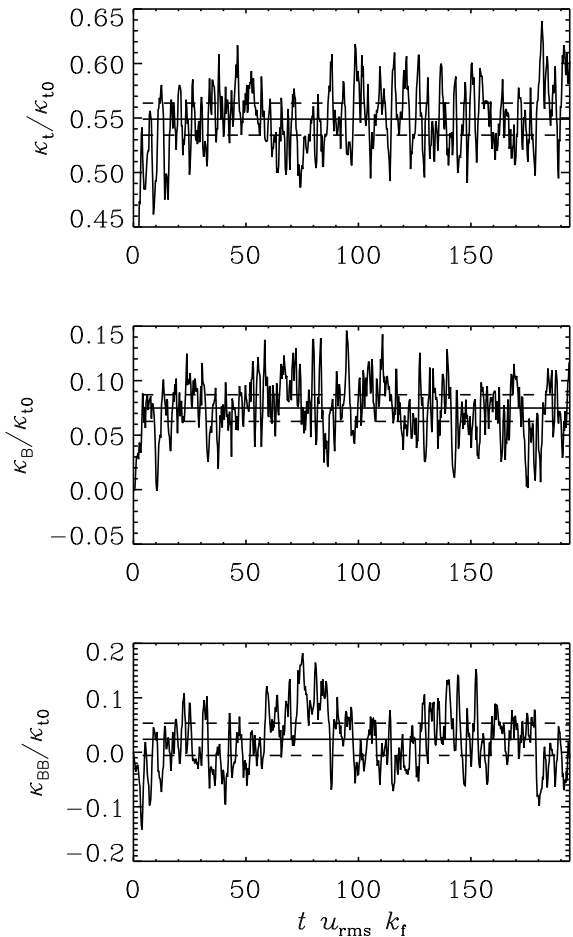}
\end{center}\caption[]{
Time series of $\kappat$, $\kappaO$, and $\kappaOO$ for a run
in \Fig{p32d} with $k=0$, $\kf/k_1=8$, $\Co=0.6$, and $\Pe=70$.
The average values are shown as straight solid lines and the
error margin is indicated by dashed lines.
}\label{pxz0_128d_Om01_k0}\end{figure}

In order to illustrate the meaning of the error bars, we show in
\Fig{pxz0_128d_Om01_k0} time series of $\kappat$, $\kappaO$, and
$\kappaOO$ for a run in \Fig{p32d} with $k=0$, $\kf/k_1=8$, $\Co=0.6$,
and $\Pe=70$.
Time is given in turnover times and covers nearly $200$ in those units.
This results in nearly 200 independent measurements, whose standard
deviation yields the one-sigma probability range of finding measurements
to lie in that interval.

\begin{figure}\begin{center}
\includegraphics[width=\columnwidth]{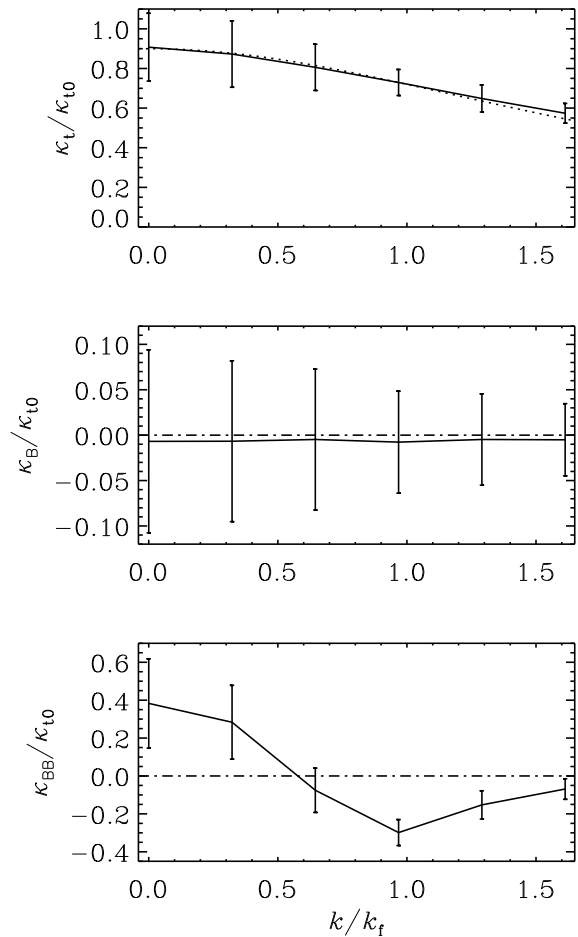}
\end{center}\caption[]{
Wavenumber dependence of $\kappat$, $\kappaB$, and
$\kappaBB$ for $\Co=0$, $\BBeq=1.4$, and $\Pe=12$.
The dotted line represents a curve of the form \eq{Lorentzian}
with $a=0.50$.
For these runs the anisotropy parameter $\varepsilon$ is around 0.13.
}\label{p32e}\end{figure}

In \Fig{p32e} we show the $k$ dependence of the transport coefficients
in the presence of a magnetic field and without rotation.
The $k$ dependence of $\kappat$ is about that found in \Fig{p32d},
but now $\kappaBB$ shows a marked dependence with a sign change
at $k/\kf\approx0.5$ about values of $\kappaBB$ of opposite sign
with extrema of similar modulus, $\kappaBB/\kappatz\approx+0.3$ at
$k/\kf\approx0$ and $\kappaBB/\kappatz\approx-0.3$ at $k/\kf\approx1$.
Furthermore, for $k\to0$ and $k\to\infty$ the values of $\kappaBB$
are small.

\section{Conclusions}

The test scalar formalism described above proves to be a useful tool
for calculating the components of the turbulent diffusion tensor for
the mean passive scalar concentration.
Here the mean passive scalar concentration is defined as a two-dimensional
field obtained by averaging over one coordinate direction.
Therefore the diffusion tensor has 6 relevant components.
In the present investigation, we have only looked at the case of
homogeneous turbulence where anisotropy is produced either by rotation
or by a magnetic field.
Another interesting alternative would be to look at the effects of a
linear shear flows of the form $\meanUU^{S}=(0,Sx,0)$.
Investigations of the corresponding magnetic case with the test field
method have already been carried out (Brandenburg et al.\ 2008a),
and they confirmed the similarity between cases with shear and rotation.
The application of the test scalar method to cases with shear would
therefore be straightforward.

The test scalar method is analogous to the test field method for obtaining
turbulent transport coefficients for the evolution of the mean magnetic
field.
Therefore, many of the questions raised in that case carry over to the
present case.
An example is the dependence of the transport coefficients on the length
and time scales of the mean field.
Usually, rapid spatial and temporal variations are less efficiently diffused
by turbulence whose correlation length and time are no longer small compared
with the corresponding scale of variation of the mean field.

A surprising result of the present work is the fact that turbulent transport
along the rotation axis is not strongly reduced unless the length scale of the
mean field becomes comparable with the correlation length of the turbulence,
i.e.\ when $k\approx\kf$.
In that case, there is significant suppression of the turbulence along the
direction of rotation.
A similar result also applies if anisotropy is produced by an applied
magnetic field.

The present investigation is restricted to the case when the scalar field
is passive.
This applies to the evolution of the entropy density in situations where
an isothermal equation of state applies and where
gravitational stratification is negligible.
However, once gravitational stratification or non-isothermal effects
become important, the
test scalar formalism can only be used to describe small departures
from a given reference state.  It would be worthwhile exploring this case further.

Our work has shown that for non-helically forced isotropic turbulence, i.e.\ without
rotation or magnetic fields, the turbulent diffusivity of a passive scalar
is similar to the value of the turbulent diffusivity of the magnetic field
in the kinematic limit.
While this was expected based on results from the first-order
smoothing approximation, the situation is less clear for turbulent
kinematic viscosity, which suggest that the turbulent magnetic Prandtl
number should be 2/5.
Simulations of Yousef et al.\ (2003), based on the decay rates of
large-scale magnetic and velocity structures, have shown that the
turbulent magnetic Prandtl number is indeed also unity.
However, their method is not very accurate and it would be good to
reconsider this by developing a method analogous to the test field and
test scalar methods for test flows.

\section*{Acknowledgments}
We thank G\"unther R\"udiger for detailed discussions regarding the sign
of $\kappaO$, and for telling us about a sign error in the corresponding
expression in Kitchatinov et al.\ (1994).
The computations have been carried out at the National Supercomputer
Centre in Link\"oping and at the Center for Parallel Computers at the
Royal Institute of Technology in Sweden.
This work was supported in part by the Swedish Research Council,
grant 621-2007-4064 (AB)
and the Sweden America Foundation through the Borgr\"attsfonderna, administered by the Office of Marshall of the Realm of Sweden (AS).

\appendix
\section{The positive sign of $\kappaO$}
\label{KappaOmega}

In this appendix we motivate the origin of the positive sign of
$\kappaO$ using the pressureless $\tau$ approximation.
In this approach one considers the evolution equation for $\meanFFFF$,
\EQ
{\partial\meanF_i\over\partial t}
=\overline{\dot{u}_i c}+\overline{u_i\dot{c}},
\EN
where dots denote time derivatives (Blackman \& Field 2003,
Brandenburg et al.\ 2004).
We use \Eqs{dUdt}{dcdt}, assume that $\uu=\UU$, i.e.\ that $\meanUU=0$,
ignore the pressure term and magnetic fields, assume $\overline{\ff c}=0$,
and obtain
\EQ
\overline{\dot{u}_i c}=-2\Omega_j\epsilon_{ijk}\overline{u_kc}
+\mbox{triple correlations},
\EN
\EQ
\overline{u_i \dot{c}}=-\overline{u_iu_j}\,\nabla_j\meanC
+\mbox{triple correlations}.
\EN
The triple correlation terms result from the nonlinearities in the
evolution equations \Eqs{dUdt}{dcdt}.
In the $\tau$ approximation we substitute the sum of the triple
correlations by quadratic correlations, i.e.\ by $-\overline{u_ic}/\tau$
in the present case (Vainshtein \& Kitchatinov 1983, Kleeorin et al.\ 1990).
We write the resulting equation in matrix form,
\EQ
\tau{\partial\meanF_i\over\partial t}=-\LLLL_{ik}\meanF_k
-\tau\overline{u_iu_j}\,\nabla_j\meanC,
\EN
where $\LLLL_{ik}=\delta_{ik}+2\Omega_j\tau\epsilon_{ijk}$.
We solve this equation for $\meanFFFF$ and obtain
\EQ
\meanF_i=-(\LLLL^{-1})_{ij}\left(\tau\overline{u_ju_k}\,
\nabla_k\meanC+\tau{\partial\meanF_i\over\partial t}\right),
\EN
where
\EQ
\LLLL^{-1}={1\over1+\Co^2}
\pmatrix{1 & \Co & 0 \cr -\Co & 1 & 0 \cr 0 & 0 & 1+\Co^2 }
\EN
and
\EQ
\overline{u_ju_k}={\rm diag}\left(\overline{u_\perp^2},
\overline{u_\perp^2}, \overline{u_\parallel^2}\right).
\EN
In the stationary state we may ignore the time derivative
and recover \Eqs{meanF}{kappaij} with
\EQ
\kappat={\tau\overline{u_\perp^2}\over1+\Co^2},\quad
{\kappaO\over\kappat}=\Co,\quad
{\kappaOO\over\kappat}={\Co^2\!+\!\varepsilon\over1-\varepsilon},
\label{kappaOO}
\EN
where $\varepsilon=1-\overline{u_\perp^2}/\overline{u_\parallel^2}$
was defined in \Eq{varepsilon}.
In the cases reported here the value of $\varepsilon$ was never found
to be negative.
Note also that $1-\varepsilon=\overline{u_\perp^2}/\overline{u_\parallel^2}$
in the denominator of \Eq{kappaOO} is always positive.
Therefore, in addition to $\kappaO$ being positive,
also $\kappaOO$ should be positive.

\label{lastpage}

\begin{thebibliography}{}

\bibitem[]{1030}
Blackman, E. G., \& Field, G. B.\ypf{2003}{15}{L73}

\bibitem[]{1031}
Brandenburg, A.\yan{2005}{326}{787}

\bibitem[]{1032}
Brandenburg, A., K\"apyl\"a, P., \& Mohammed, A.\ypf{2004}{16}{1020}

\bibitem[]{1033}
Brandenburg, A., R\"adler, K.-H., Rheinhardt, M.,
\& K\"apyl\"a, P. J.\yapj{2008a}{676}{740}

\bibitem[]{1034}
Brandenburg, A., R\"adler, K.-H., \& Schrinner, M.\yana{2008b}{482}{739}

\bibitem[]{1035}
Durney, B. R., Roxburgh, I. W.\ysph{1971}{16}{3}{20}

\bibitem[]{1036a}
Gough, D. O.\yproc{1978}{337}
{Proc. EPS Workshop on Solar Rotation}
{G. Belvedere \& L. Patern\`o}{Catania University Press, Catania}

\bibitem[]{1036}
Haugen, N. E. L., Brandenburg, A., \& Dobler, W.\ypre{2004}{70}{016308}

\bibitem[]{1037}
Hubbard, A., \& Brandenburg, A.\sapj{2008}, arXiv:0811.2561

\bibitem[]{1038}
Julien, K., Legg, S., McWilliams, J., \& Werne, J.\yjfm{1996}{322}{243}

\bibitem[]{1039}
K\"apyl\"a, P. J., Korpi, M. J., \& Tuominen, I.\yana{2004}{422}{793}

\bibitem[]{1040}
Kitchatinov, L. L., \& Mazur, M. V.\ysph{2000}{191}{325}

\bibitem[]{1041}
Kitchatinov, L. L., R\"udiger, G., \& Pipin, V. V.\yan{1994}{315}{157}

\bibitem[]{1042}
Kitchatinov, L. L. \& R\"udiger, G.\yana{1995}{299}{446}

\bibitem[]{1043}
Kleeorin, N. I., Rogachevskii, I. V., Ruzmaikin, A. A.\yjetp{1990}{70}{878}

\bibitem[]{1044a}
Mitra, D., K\"apyl\"a, P. J., Tavakol, R., \& Brandenburg, A.\yana{2009}{495}{1}

\bibitem[]{1044}
Rieutord, M., Brandenburg, A., Mangeney, A., \&
Drossart, P.\yana{1994}{286}{471}

\bibitem[]{1045}
R\"udiger, G.\ybook{1989}{Differential rotation and stellar convection:
Sun and solar-type stars}{Gordon \& Breach, New York}

\bibitem[]{1046}
R\"udiger, G., \& Hollerbach, R.\ybook{2004}{The magnetic universe}
{Wiley-VCH, Weinheim}

\bibitem[]{1047}
R\"udiger, G., \& Pipin, V. V.\yana{2001}{375}{149}

\bibitem[]{1048}
R\"udiger, G., Egorov, P., Kitchatinov, L. L., \& K\"uker, M.\yana{2005}{431}{345}

\bibitem[]{1049}
Schrinner, M., R\"adler, K.-H., Schmitt, D., Rheinhardt, M.,
Christensen, U.\yan{2005}{326}{245}

\bibitem[]{1050}
Schrinner, M., R\"adler, K.-H., Schmitt, D., Rheinhardt, M.,
Christensen, U. R.\ygafd{2007}{101}{81}

\bibitem[]{1051}
Sur, S., Subramanian, K., \& Brandenburg, A.\ymn{2007}{376}{1238}

\bibitem[]{1052}
Sur, S., Brandenburg, A., \& Subramanian, K.\ymn{2008}{385}{L15}

\bibitem[]{1053}
Thompson, M. J., Christensen-Dalsgaard, J., Miesch, M. S.,
\& Toomre, J.\yaraa{2003}{41}{599}

\bibitem[]{1054}
Vainshtein, S. I. \& Kitchatinov, L. L.\ygafd{1983}{24}{273}

\bibitem[]{1055}
Vitense, E.\yjour{1953}{Z.\ Astrophys.}{32}{135}

\bibitem[]{1056}
Weiss, N. O.\yjour{1965}{Observatory}{85}{37}

\bibitem[]{1057}
Yousef, T. A., Brandenburg, A., \& R\"udiger, G.\yana{2003}{411}{321}

\end{thebibliography}
\end{document}